\documentclass[prl, twocolumn, floatfix,showpacs]{revtex4-1}

\usepackage{graphicx}
\usepackage{dcolumn}
\usepackage{bm}
\usepackage{enumerate}
\usepackage[latin1]{inputenc}
\usepackage{color}

\begin{document}

\preprint{APS/123-QED}

\title{Ultrahigh Bandwidth Spin Noise Spectroscopy:\\
Detection of Large g-Factor Fluctuations in Highly n-Doped GaAs}

\author{Fabian Berski}
\email{Berski@nano.uni-hannover.de}
\author{Hendrik Kuhn}
\thanks{The authors F. B. and H. K. contributed equally to this work.}
\author{Jan G. Lonnemann}
\author{Jens H\"ubner}
\email{jhuebner@nano.uni-hannover.de}
\author{Michael Oestreich}
\email{oest@nano.uni-hannover.de}
\affiliation{Institut f{\"u}r Festk\"orperphysik, Leibniz Universit\"at Hannover, Appelstr.~2, D-30167 Hannover, Germany} 
\date{\today}

\begin{abstract}
We advance all optical spin noise spectroscopy (SNS) in semiconductors to detection bandwidths of several hundred gigahertz by employing an ingenious scheme of pulse trains from ultrafast laser oscillators as an optical probe. The ultrafast SNS technique avoids the need for optical pumping and enables nearly perturbation free measurements of extremely short spin dephasing times. We employ the technique to highly n-doped bulk GaAs where magnetic field dependent measurements show unexpected large g-factor fluctuations. Calculations suggest that such large g-factor fluctuations do not necessarily result from extrinsic sample variations but are intrinsically present in every doped semiconductor due to the stochastic nature of the dopant distribution.
\end{abstract}

\pacs{72.25.Rb, 72.70.+m, 78.47.db, 85.75.-d}
\maketitle

Spin noise spectroscopy (SNS) has proven itself as a well-developed experimental technique in semiconductor spin quantum-optronics \cite{Muller.PhysicaE.2010, Dahbashi.APL.2012, Crooker.PRB.2009}. The low perturbing nature of SNS makes the technique an ideal tool to study the unaltered long coherence times of electron spins in semiconductors \cite{Romer.PRB.2010} and semiconductor nanostructures \cite{Muller.PRL.2008, Dahbashi.APL.2012}. However, short spin coherence times require a high detection bandwidth and thus the temporal capabilities of SNS are usually limited by the speed of the electro-optic conversion and subsequent signal processing. A first successful step to overcome the temporal limitation has been made by employing a single ultrafast laser oscillator as a stroboscopic optical sampling tool, which directly enabled spin noise measurements of frequencies up to several GHz, but with a fixed bandwidth of roughly 0.1 GHz \cite{Muller.PRB.2010}. In this letter we report the first experimental demonstration of spin noise spectroscopy with a full bandwidth that is increased by several orders of magnitude to over one hundred GHz which corresponds to spin dephasing times in the picosecond regime. Thereby, the presented SNS method is ideally suited for systems which intrinsically show a fast decay of spin coherence and are yet susceptible to optical excitation, like hole spin systems with a high degree of spin-orbit interaction \cite{Wu.PhysicsReports.2010}, carrier systems at very low temperatures ($<$100 mK), or Bose-Einstein condensation of magnons \cite{Demokritov.Nature.2006}.

In the following, we employ the technique of ultrafast SNS to highly n-doped bulk GaAs and find in the metallic regime large g-factor fluctuations. Calculations reveal that large g-factor fluctuations are an intrinsic bulk property of doped semiconductors. The effect results from the stochastic distribution of donor atoms and the imperfect local averaging of electrons due to their finite momentum and spin dephasing times \cite{Dzhioev.PRB.2002}. Ultrafast SNS asserts itself as the perfect tool to measure such an effect since it combines the necessary high temporal resolution, negligible disturbance of the system, and efficient averaging over very large sample volumes compared to other optical methods due to the below-bandgap detection. 

\begin{figure}[hb]
  \centering
  \includegraphics[width=0.95 \columnwidth]{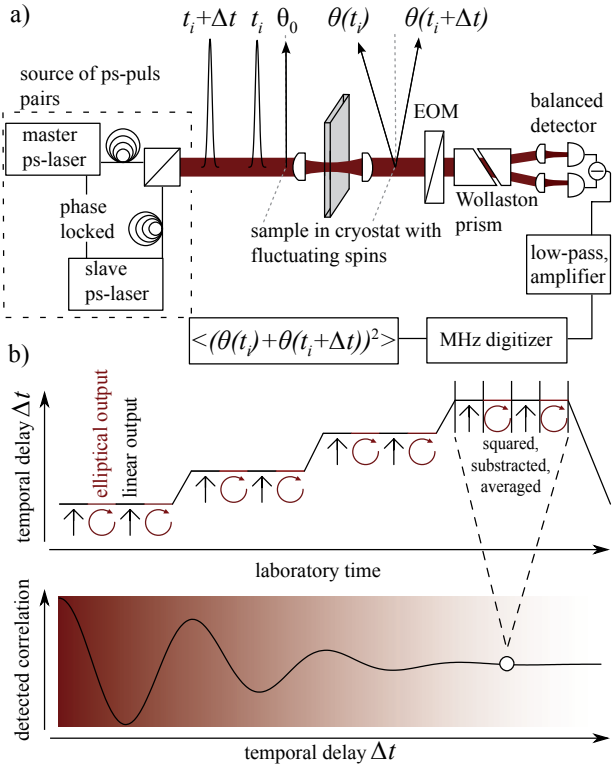}
  \caption{(Color online) a) Experimental Setup: The outputs of two synchronized ps-laser oscillators are combined via a single mode optical fiber and transmitted through the sample (grey) which is mounted in a magneto-optical cryostat (not shown). The rotation of the linear input polarization $\theta_{0}$ is analyzed behind the sample with a polarization bridge and a balanced photoreceiver. Here, $\theta(t_{i})$ and $\theta(t_{i}+\Delta t)$ denote the stochastic rotation of the two individual laser pulses after the sample. The amplified electrical signal is digitized and seamlessly analyzed on a standard computer. b) Schematic measurement sequence. Details are given in the text. 
  }
\label{fig:Aufbau}
\end{figure}

The extended measurement principle of ultrafast SNS is based upon the repeated measurement of the correlated Faraday rotation signal $ \theta(t_{i}) \theta(t_{i}+\Delta t)  $ of two ultrashort laser probe pulses with a temporal delay of $\Delta t$ \cite{Starosielec.APL.2008}. The point in time $t_{i}$ is arbitrary for every pulse pair due to the stochastic nature of the spin dynamics if the repetition period between two pulse pairs is much larger than the spin dephasing time. The average Faraday rotation signal $\left\langle \theta \right\rangle (\Delta t)$ vanishes if the non-magnetic sample is in thermal equilibrium. However, the variance $ \sigma _{\theta }^{2}  (\Delta t)$ is not zero but is maximal for fully correlated Faraday rotation of the two laser pulses ($\Delta t=0$), decreases with increasing $\Delta t$ to a finite value due to spin dephasing, oscillates with $\Delta t$ in the presence of a transverse magnetic field $B$ due to Larmor precession of the electron spins, and approaches zero for anti-correlation. 

In this work the sampling pulses are delivered by two synchronized, ultrafast, picosecond laser oscillators with a common repetition rate of 80~MHz. The relative phase between the two emitted pulse trains is adjustable, so that pairs consisting of two laser pulses are formed with a temporal delay $\Delta t$ which can be conveniently tuned between a picosecond and a few nanoseconds. The correlated Faraday rotation signal of both pulses within a pulse pair is measured by a balanced detector which is so slow that it integrates over each pulse pair but is fast enough to distinguish two succeeding pulse pairs. In other words, the Faraday rotation signals $\theta (t)$ of the two pulses of a pulse pair are added up for $\Delta t < 12.5$~ns but the fluctuation from pulse pair to pulse pair is fully resolved. A rectification of the Faraday signal is implemented by taking the square of $\theta (t_i)+\theta (t_i+\Delta t)$ during the data acquisition. The experimental setup is depicted in Fig.~\ref{fig:Aufbau}. The two degenerate, linearly polarized laser pulses are combined in a polarization maintaining, single mode fiber to ensure a common beam profile in addition to identical pulse length, power, and wavelength. The  blended laser light has an average power of 17~mW and is focussed to a spot diameter of about 50~$\mu$m onto the sample surface. After traversing the sample, the spin induced fluctuations of the linear polarization are analyzed by a polarization bridge given by a $1/2 \cdot \lambda $ waveplate for power balancing, a Wollaston prism, and a low noise, differential, optical photoreceiver with a 3~dB bandwidth of 150~MHz. The electrical output of the photoreceiver is passed through a low pass filter with a cut-off frequency of 70~MHz before being amplified in order to suppress any residual voltage peaks arising from the limited common noise rejection of the differential photoreceiver. Finally the filtered signal is digitized by a 180~MSample/s digitizer card and sent to a PC for further processing. 

The measured signal is not only composed of pure spin noise but also of residual background contributions, which are mainly caused by optical shot noise. The spin noise is extracted by using an electro optical modulator (EOM) before the polarization bridge, which acts either as $1/4 \cdot \lambda $ or as $0\cdot \lambda $ retarder with a square wave modulation of 4~kHz. For $0\cdot \lambda $ retardance, the EOM transmits the incoming polarization unchanged (spin noise is detected), but for $1/4\cdot \lambda $ retardance, every off-axes polarization component is transformed into elliptically polarized light and divided into two equal parts at the polarization bridge (no spin noise is detected; background only). This fast background acquisition strongly suppresses any parasitic fluctuations and yields extremely reliable data series. The measurement protocol is depicted in Fig.~\ref{fig:Aufbau}b: A single measurement window is 100~ms long. The start and end point are set in the presented measurement to 80~ps and 835~ps, respectively. The time delay is increased in 96 steps with step length of 1~ms. The exact time delay for each step has been verified with a calibrated streak camera system. During each step the EOM switches eight times between $1/4 \cdot \lambda $ and $0 \cdot \lambda $ retardance. 

\begin{figure}[t]
  \centering
  \includegraphics[width=0.99 \columnwidth]{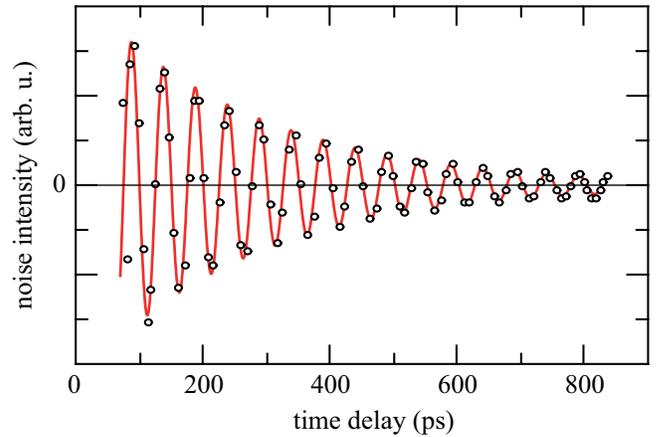}
  \caption{(Color online) Spin correlation of free electron spins precessing in a transverse magnetic field of 6 T. The measurement is taken at 20~K with a probe laser photon energy detuned by $\approx 40$~meV below the absorption edge (dots) and the red line is a fit to the data according to Eq.~\ref{eq:ddtautocorr}. }
\label{fig:spectrum}
\end{figure}

The sample is Czochralski grown, bulk GaAs with a nominal n-doping concentration of $n_{d} = 8.2\cdot 10^{17}$~cm$^{-3} $. The doping concentration is about 40 times above the metal-to-insulator transition which guarantees that all valence electrons are in the metallic state. In lower doped samples, localized and free electrons can exist simultaneously and strong g-factor variations are consequently trivial. The sample thickness is $300\,\mu$m and both front and back surfaces are anti-reflection coated for optimized transmission. The high doping concentration yields a Fermi-level of $E_{F} =47.7$~meV above the conduction band minimum. The dominating spin dephasing mechanism in this regime is the Dyakonov-Perel \cite{Dyakonov.SPSS.1972} mechanism since the resulting energy dependent spin-splitting of the conduction band is large. Accordingly, we expect a very fast spin dephasing time on the order of some hundred picoseconds \cite{Dzhioev.PRB.2002}.

Figure~\ref{fig:spectrum} shows the derivative of the measured (dots) spin correlation ${\frac{d}{dt}}  \sigma _{\theta }^{2} $ as a function of the temporal pulse delay $\Delta t$. The derivative has been taken to suppress a slow varying background slope with $\Delta t$ which originates from the coupling of the two independent laser sources by a lock-to-clock system \footnote{A free running ultrafast rapid temporal delay scanning scheme circumvents this background but raises other constraints which is discussed in detail in Ref. \cite{Hubner2012}.}. Assuming a free induction decay of the free precessing electrons, $\cos(\omega_{L}t') e^{-t'/\tau_{s}}$, the derivative of the autocorrelation is given by:
\begin{equation}
\frac{d}{dt}  \sigma _{\theta }^{2}  \propto \left\{\omega _{L} \tau _{s} \sin (\omega _{L} t)+2\cos (\omega _{L} t)\right\}e^{-t/\tau _{s} } \label{eq:ddtautocorr}
\end{equation}
where $\omega _{L} =\hbar ^{-1} g^{*} B$ is the Larmor precession frequency with $g^{*} $ as effective electron g-factor and $\tau _{s} $ is the spin dephasing, i.e., spin correlation time. The red line in Fig.~\ref{fig:spectrum} is a fit with Eq.~(\ref{eq:ddtautocorr}) which matches with very high accuracy.

\begin{figure}[t]
  \centering
  \includegraphics[width=0.99 \columnwidth]{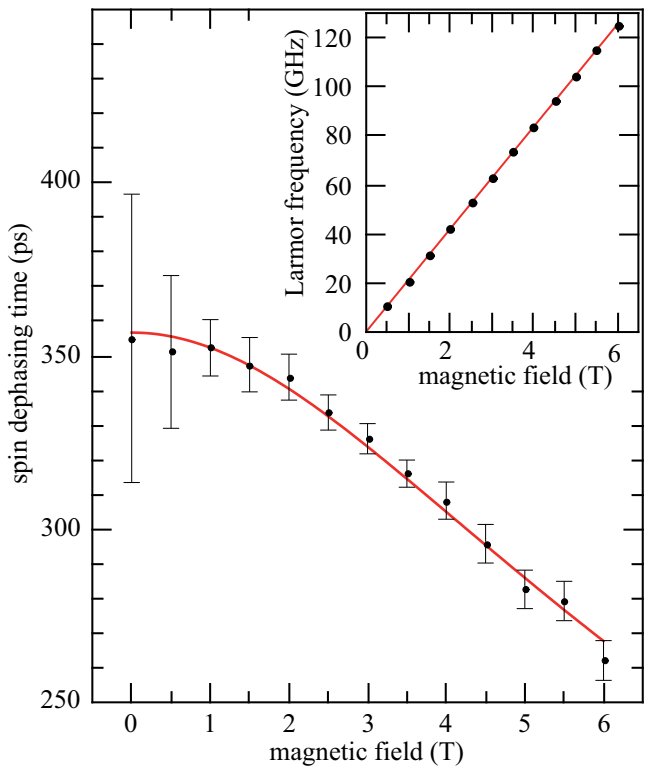}
  \caption{(Color online) Dependence of spin dephasing time $\tau _{s} $ on the transverse magnetic field strength measured at a sample temperature of 20 K. The inset shows the measured change of the Larmor frequency $\omega _{L} $ with magnetic field (black dots) and a linear fit through the origin (red line) which yields the electron Land{\'e} g-factor.
  }
\label{fig:B-Dependence}
\end{figure}

Figure~\ref{fig:B-Dependence} shows the extracted dependence of spin dephasing time on the applied transverse magnetic field strength $B$. The spin dephasing time of $\tau _{s} \approx 360$~ps at vanishing magnetic field corresponds well to the expected spin dephasing time limited by the Dyakonov-Perel spin dephasing mechanism for the nominal doping concentration of the investigated sample \cite{Dzhioev.PRB.2002}. Surprisingly, the spin dephasing time decreases with increasing magnetic field due to a significant inhomogeneous spread of the electron Land{\'e} g-factor. The red line in Fig.~\ref{fig:B-Dependence} is a fit given by the inverse width $w_{v}$ of an approximated Voigt profile according to:
\begin{equation}
\tau_{s} =\left(\pi \, w _{v} \right)^{-1} \approx \left(c_{0} \gamma _{h} +\sqrt{c_{1} \gamma _{h}^{2} +\gamma _{i}^{2} } \right)^{-1} \label{eq:voigt}
\end{equation}
where $\gamma _{h} ,\gamma _{i} $ are the homogenous and inhomogeneous spin dephasing rates, respectively. The factor $\pi $ arises from the width of the Fourier transformation of a mono-exponential decay which we adopt as a valid approximation for the data analysis \cite{Romer.RSI.2007}; $c_{0} $ and $c_{1} $ are constants \footnote{$c_{0} ={\rm 0.5346},c_{1} =0.2166$ \cite{Olivero.JQSRT.1977}}. The inhomogeneous spin dephasing rate $\gamma _{i} $ is directly linked to the standard deviation of the g-factor spread $\sigma _{g} $ by $\gamma _{i} =\sigma _{g} \mu _{B}  B/\hbar $. From the fit with Eq.~(\ref{eq:voigt}) we obtain a g-factor variation $\sigma _{g} =0.0032$ which is surprisingly large taking into account that all valence electrons are well in the metallic state. We will explain the possible origin of this phenomena in the next paragraph. The inset of Fig.~\ref{fig:B-Dependence} depicts the dependence of the Larmor precession frequency on $B$. The relative measurement error of the Larmor frequency is smaller than $10^{-4}$ for $B \ge 2$~T while the absolute error is about $\pm 1$~\% due to errors in the absolute calibration of $B$ and $\Delta t$. Please note that the demonstrated full bandwidth of about 120~GHz is only limited by the highest magnetic field of 6~T. The nearly perfect fit to a straight line yields the magnitude of the average free electron Land{\'e} g-factor which is $g^{*} =-0.241$. The negative sign is assigned from the relation $g^{*} =-0.48+\beta \cdot E_{F} $. We determine the factor $\beta $ which reflects the energy dependence of the Land{\'e} g-factor to $\beta \approx 5.1$~eV$^{-1}$ for this doping concentration and attribute the deviation from the commonly known factor of $6.3$~eV$^{-1}$ for slightly doped samples \cite{Hopkins.SST.1987,Hubner.PRB.2009} to band gap renormalization arising from the high doping concentration. 
The deviation of $g^{\ast}$ being a constant is less than $10^{-3}$~T$^{-1}$ which is at least a factor of 5 lower than for low doped GaAs at low temperatures. 

Next, we discuss the origin of $\sigma_g$.
Most interestingly, the measured g-factor variation in metallic bulk semiconductors can be attributed to an intrinsic contribution which arises from the pure thermodynamic distribution of dopant atoms in the material during growth. The Fermi--level is inherently constant over the entire sample, but the stochastic fluctuations of the dopant concentration give rise to local space charge densities \cite{PhysRevB.81.115332} which in turn shift the band structure with respect to the Fermi--level. In first approximation, an electron propagates in this local inhomogeneity undisturbed over an average distance $\overline r = v_f \cdot \tau_p /2$, where $\tau_p$ is the electron momentum scattering time and $v_f = \sqrt{2 E_F / m^*}$ is the Fermi--velocity.  At low temperatures ionized impurity scattering is the main scattering mechanism for highly doped bulk semiconductors. The momentum scattering time can easily be extracted from the spin dephasing time measured at zero magnetic field by the relation \cite{Pikus.SR.1984, Zutic.RMP.2004}:
\begin{equation}
\tau_{s}^{-1} = \frac{{32}}{{105}}{\gamma _3}^{ - 1}{}{\alpha ^2}\frac{{{E_F}^3}}{{{\hbar ^2}E_g}} \cdot \tau _p \label{eq:taus}
\end{equation}
with $\gamma_3=6$ for ionized impurity scattering and $\alpha = 0.07$ \cite{Marushchak.SPSS.1983}; $E_{g}$ is the energy gap. We determine from the measured $\tau_{s} \approx 360$~ps an average momentum scattering time of 70~fs, which is very reasonable for this kind of sample and scattering mechanism \cite{Zutic.RMP.2004}. An electron samples an average volume $\overline{V}=4 \pi /3   \, \overline{r}^3 \cdot \tau_s/ \tau_p $ during $\tau_{s}$ on its diffusive scattering path. For each electron the number of donor atoms within this volume fluctuates with $\sqrt{\overline{V} \cdot n_{d}}$. The resulting change in the local doping density is directly linked to the g-factor variation via the energy dependence of the g-factor as shown above. For the given doping density in our sample we calculate an intrinsic g-factor variation due to the local doping density fluctuations of $\sigma_{g} = 5\cdot 10^{-4}$. The experimental value is only a factor of six higher than this calculated value from our stochastic approximation.

Certainly, other inhomogeneities may contribute to the measured g-factor fluctuation and the theoretical description is only an order of magnitude estimation. Nevertheless, a statistical distribution of donor atoms is inevitably present in doped semiconductor samples and the estimated effect on $\sigma_g$ is large. The effect should be in particular orders of magnitude larger than the familiar variable g-factor mechanism due to electrons in different quantum states \cite{PhysRevB.66.233206}. We want to point out that our estimation fully links $\sigma_g$ to the doping density since the spin dephasing rate is related to the momentum scattering rate by Eq.~\ref{eq:taus} and the momentum scattering rate is in turn related to the doping density by the Brooks-Herring formalism. The Fermi-level and Fermi-velocity are by definition determined by $n_{\rm d}$ and, therefore, the intrinsic inhomogeneous g-factor fluctuation should scale according to this approximation like $\sigma_{g}\propto n_{\rm d}^{2/3}$. The calculations also predict a larger $\sigma_g$ for higher temperatures since the faster spin and momentum relaxation times yield a smaller averaging volume. We measured such an increase of $\sigma_g$ by a factor of two for a temperature increase from 20~K to 200~K. However, we also measured a lateral g-factor fluctuation of about $\sigma _{g}^{{\rm lat}} \approx 0.0015$ by spatially changing the transmission spot on the sample and our current measurement setup does not guarantee a constant sampling spot if the temperature is increased. Clearly, further measurements with improved spot stability and doping dependent measurements on thick, high quality, molecular beam epitaxy grown GaAs together with sophisticated microscopic calculations are desirable to test and quantify the effect. 

In conclusion we successfully demonstrated ultrafast spin noise spectroscopy and increased the state of the art bandwidth by more than two orders of magnitude. The bandwidth is in principle only limited by the pulse width of the laser and should reach for femtosecond laser pulses the THz regime \footnote{Without loss of applicability, one ultrafast laser together with a mechanical delay stage can be used instead of two ultrafast lasers.}. Already the demonstrated bandwidth of 120~GHz enables SNS measurements on systems with picosecond spin dynamics. This applies for, e.g., magnons in yttrium iron garnet, hole spin systems at very low temperatures, as well as for many-electron systems at room temperature. We applied ultrafast SNS to highly n-doped bulk GaAs well above the metal-to-insulator transition, which is the archetype material for spintronic, and observed, despite being in the metallic regime, a large g-factor variance. Calculations estimate that such large g-factor variances are intrinsic to doped semiconductors and result even in perfect samples from the inevitable stochastic variation of the doping concentration.

We acknowledge the financial support by the BMBF joint research project QuaHL-Rep, the Deutsche Forschungsgemeinschaft in the framework of the priority program ``SPP 1285---Semiconductor Spintronics,'' and the excellence cluster ``QUEST---Center for Quantum Engineering and Space-Time Research''.

\end{document}